\documentclass[pra]{revtex4}
\usepackage{amsmath}
\usepackage{graphicx}
\begin{document}
\title{Mechanical Motion Induced by Spatially Distributed Limit-Cycle Oscillators}
\author{Hidetsugu Sakaguchi and Yuuki Mukae}
\affiliation{Department of Applied Science for Electronics and Materials,
Interdisciplinary Graduate School of Engineering Sciences, Kyushu
University, Kasuga, Fukuoka 816-8580, Japan}
\begin{abstract}
Spatially distributed limited-cycle oscillators are seen in various physical and biological systems. 
 In internal organs, mechanical motions are induced by the stimuli of spatially distributed limit-cycle oscillators. 
We study several mechanical motions by limit-cycle oscillators using simple model equations. One problem is deformation waves of radius oscillation induced by desynchronized limit-cycle oscillators, which is motivated by peristaltic motion of the small intestine. A resonance-like phenomenon is found in the deformation waves, and particles can be transported by the deformation waves. Another is the beating motion of the heart. The expansion and contraction motion is realized  by a spatially synchronized limit-cycle oscillation; however, the strong beating disappears by spiral chaos, which is closely related to serious arrhythmia in the heart. 
\end{abstract}
\maketitle
\section{Introduction}
 Rhythmic motions and mutual synchronization have been intensively studied~\cite{Winfree,Kuramoto,Pikovsky}. One  important topic is  mutual synchronization in spatially distributed limit-cycle oscillators~\cite{Saka}. There are various problems in which mechanical  motions are induced by spatially distributed oscillators.  
For example, walking rhythms are induced by rhythmic stimuli from the brain~\cite{Ijspeert}. Some internal organs exhibit rhythmic motion. The heart repeats expansion and contraction, which is induced by the rhythmic electrophysiological activity of heart cells~\cite{Negroni}. Similarly, the small intestine exhibits peristaltic motion according to the electrophysiological rhythm of intestinal cells~\cite{Davenport}. 

The mechanical deformation of internal organs can be described as viscoelastic materials. There are some numerical studies using realistic but very complicated model equations~\cite{Zajac,Ijiri}. We would like to understand qualitatively 
the relationship between basic rhythms and induced mechanical motions from the viewpoint of nonlinear physics.  
In this paper, we will study mechanical motions by spatially distributed limit-cycle oscillators using simple model equations. Furthermore, we will consider particle motions driven by deformation waves, which is analogous to the transport of food particles by the peristaltic motion of the small intestine.  In Sect. 2, we consider simple elastic waves induced by sinusoidal waves. We also consider breathing motions (radius oscillations) driven by sinusoidal oscillations. In Sect. 3, we consider a simple model of a particle motion in a traveling wave. By generalizing these simple and tractable models, we study mechanical motions induced by desynchronized limit-cycle oscillations and chaotic oscillations. In Sect. 4, we use the Stuart--Landau equation with a natural frequency distribution as a rhythm generator, and consider the effect of desynchronization on elastic waves. We investigate particle motions by the desynchronized deformation waves. 
In Sect. 5, we use the Aliev--Panfilov equation as a rhythm generator in a two-dimensional sheet, and consider the effect of spatiotemporal chaos on the area oscillation of the sheet. It is closely related to serious arrhythmia such as the ventricular fibrillation of the heart.  
\section{Elastic Waves Driven by Simple Sinusoidal Waves}
A simple model of an elastic wave is a coupled oscillator system:
\begin{equation}
m\frac{d^2x_j}{dt^2}=K(x_{j+1}-x_{j}-a_0)-K(x_j-x_{j-1}-a_0)-h\frac{dx_j}{dt},
\end{equation}
where $m$ is mass, $K$ is the spring constant, $h$ is the viscous coefficient, and $a_0$ is the natural length of the spring. If the natural length $a_0$ of a spring changes in time as a sinusoidal wave, the equation of motion is expressed as
\begin{eqnarray}
m\frac{d^2x_j}{dt^2}&=&K[x_{j+1}-x_{j}-a_0-a_1\cos(kja_0+ka_0/2-\omega t)]\nonumber\\
& &-K[x_j-x_{j-1}-a_0-a_1\cos(kja_0-ka_0/2-\omega t)]-h\frac{dx_j}{dt},\end{eqnarray}
where $a_1$ is the amplitude of the sinusoidal wave and $k$ is the wavenumber.  This equation has a form of coupled forced oscillations. In an infinite system, the solution is expressed as $x_j(t)=A\sin(kja_0-\omega t+\delta)$, where $A$ and $\delta$ satisfy
\begin{equation}
A=\frac{2Ka_1\sin(ka_0/2)}{\sqrt{\{2K[(1-\cos(ka_0)]-m\omega^2\}^2+\omega^2h^2}},\;\tan\delta=\frac{\omega h}{[2K(1-\cos(ka_0)]-m\omega^2}.
\end{equation}
If $h=0$ or there is no viscous term, there is a resonance at $\omega/k=\sqrt{K/m}a_0$ for sufficiently small $k$. This condition implies that the wave velocity $\omega/k$ of the forced terms is equal to the intrinsic wave velocity $\sqrt{K/m}a_0$ of the coupled linear equations.  

For a uniform forcing or $k=0$, the amplitude $A$ becomes zero in an infinite system. However, in a finite system, the oscillation amplitude is not zero. For example, we can assume a circular system of $N$ elements as shown in Fig.~1(a) and consider a breathing motion of the radius $R(t)$. The small intestine has a tubular structure, and the motion of a circular portion in the cross section can be considered as a breathing motion.  Heart beats can  also be interpreted as such radius oscillation. In any case, the equation for $R(t)$ can be derived from Eq.~(2) by the projection of the equation of motion for $x_i(t)$ in the radial direction.
If $2\pi/N$ is expressed as $\Delta\theta$, $R(t)$ obeys
\begin{equation}
m\frac{d^2R}{dt^2}=-4K\sin^2(\Delta\theta/2)[R-R_0-R_1\cos(\omega t)]-H\frac{dR}{dt},
\end{equation}
where $R_0=a_0/[2\sin(\Delta\theta/2)]$, $R_1=a_1/[2\sin(\Delta\theta/2)]$, and $H=h/[2\sin(\Delta\theta/2)]$. 
 This is an equation of forced oscillation. 
The radius exhibits oscillation of $R(t)=R_0+A\cos(\omega t-\delta)$, where 
\[A=\frac{4K\sin^2(\Delta\theta/2)R_1}{\sqrt{[4K\sin^2(\Delta\theta/2)-m\omega^2]^2+\omega^2H^2}},\;\tan\delta=\frac{\omega H}{4K\sin^2(\Delta\theta/2)-m\omega^2}.\]        
\begin{figure}[tbp]
\begin{center}
\includegraphics[height=3.cm]{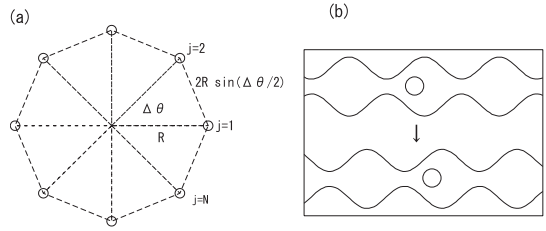}
\end{center}
\caption{(a) Schematic figure of a circular chain for Eq.~(4). (b) Schematic figure of the particle motion trapped in a traveling wave. }
\label{fig1}
\end{figure}
\begin{figure}[tbp]
\begin{center}
\includegraphics[height=4.cm]{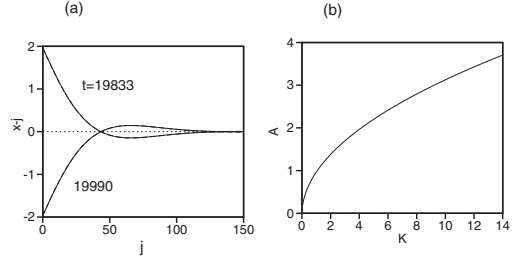}
\end{center}
\caption{(a) Profiles of the deviation $x_j(t)-ja_0$ for $0\le j \le 150$ at $t=19833$ and $t=19990$ for $a_0=1,a_1=0.1,m=1,K=4,\omega=0.02$, $h=0.5$, and $N=300$ in Eq.~(5). (b) Amplitude $A$ as a function of $K$ for $a_0=1,a_1=0.1,m=1,\omega=0.02$, and $h=0.5$.  }
\label{fig2}
\end{figure}

Next, we consider a linear system with no-flux boundary conditions.  
This is a simple model of an active filament of finite size such as a muscle fiber. 
The total length of the fiber oscillates in time owing to the expansion and contraction of the  elements. The model equation is expressed as 
\begin{eqnarray}
m\frac{d^2x_j}{dt^2}&=&K[x_{j+1}-x_{j}-a_0-a_1\cos(\omega t)]-h\frac{dx_j}{dt},\;{\rm for}\;j=0\nonumber\\ 
m\frac{d^2x_j}{dt^2}&=&K(x_{j+1}-2x_{j}+x_{j-1})-h\frac{dx_j}{dt},\;{\rm for}\; 1\le j\le N-1\nonumber\\
m\frac{d^2x_j}{dt^2}&=&-K[x_{j}-x_{j-1}-a_0-a_1\cos(\omega t)]-h\frac{dx_j}{dt},\;{\rm for}\;j=N.
\end{eqnarray}
The oscillators at $j=0$ and $N$ are coupled with only one inner neighbor.  There is no forcing term for other oscillators of $1\le j\le N-1$, because the same magnitudes of forces are applied from two neighbors and the forces are cancelled out.   
If the solution is assumed as $x_j=ja_0+Ae^{-\alpha_1 ja_0}\cos(\omega t-\alpha_2 ja_0-\delta)$ for $j<N/2$ and $x_j=ja_0+Ae^{-\alpha_1 (N-j)a_0)}\cos[\omega t-\alpha_2 (N-j)a_0-\delta]$ for $j>N/2$, $\alpha_1,\alpha_2,A$, and $\delta$ satisfy
\begin{eqnarray}
-m\omega^2A&=&K[e^{-(\alpha_1+i\alpha_2)a_0}-2+e^{(\alpha_1+i\alpha_2)a_0}]A-i\omega hA\nonumber\\
-m\omega^2A&=&K[e^{-(\alpha_1+i\alpha_2)a_0}-1]A-Ka_1e^{i\delta}-i\omega hA.
\end{eqnarray}
\begin{figure}[tbp]
\begin{center}
\includegraphics[height=4.cm]{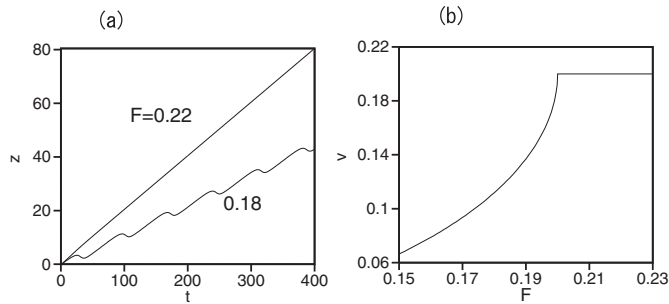}
\end{center}
\caption{(a) Time evolution of $z(t)$ at $F=0.22$ and 0.18 for $M=k=\gamma=1$, and $\omega=0.2$ in Eq.~(9). (b) Average velocity as a function of $F$ for $M=k=\gamma=1$, and $\omega=0.2$ in Eq.~(9).}
\label{fig4}
\end{figure}
Figure 2(a) shows the profiles of the deviation $x_j(t)-ja_0$ from the equilibrium position at $t=19833$ and $t=19990$ at $a_0=1,a_1=0.1,m=1,K=4,\omega=0.02$, and $h=0.5$ for $N=300$. The solutions to Eq.~(6) are $\alpha_1=0.0347,\alpha_2=0.0361,A=-1.965$, and $\delta=0.823$. Figure 2(a) shows the profiles obtained by direct numerical simulation and the solution $x_j-ja_0=Ae^{-\alpha_1 ja_0}\cos(\omega t-\alpha_2 ja_0-\delta)$. The two profiles completely overlap. The deformation is localized near the two boundaries, because force terms appear only at $j=0$ and $N$ in Eq.~(5). The total length $x_N(t)-x_0(t)\sim 2A\cos(\omega t-\delta)$ oscillates with the frequency $\omega$. Figure 2(b) shows the relationship between $K$ and $A$. The amplitude $A$ of the oscillation of the total length increases as $A\propto K^{1/2}$. 
If $a_0=1,\alpha_1<<1,\alpha_2<<1,\omega<<1,h=O(1)$, and $K=O(1)$ are assumed, the following approximate relations are derived.
\begin{equation}
\alpha_1^2-\alpha_2^2=-\frac{m\omega^2}{K},\;\alpha_1\alpha_2=\frac{\omega h}{2K},\;K^2a_1^2=[(m\omega^2-K\alpha_1)^2+(K\alpha_1+\omega h)^2]A^2.
\end{equation}
These relations and $\omega<<1$ yield 
\begin{equation}
\alpha_1=\alpha_2=\sqrt{\omega h/(2K)}, A=\sqrt{K/(\omega h)}a_1.
\end{equation}
For $a_1=0.1,\omega=0.02$, and $h=0.5$, $A=\sqrt{K}$, which is a rough approximation for the numerical solution to Eq.~(6) shown in Fig.~2(b). 
  
\section{Particle Motion in a Traveling Wave}
In this section, we consider the motion of a particle of mass $M$ induced by a traveling wave of wavenumber $k$ and frequency $\omega$. It is a very simple model of the transport of food particles by the peristaltic motion of the small intestine as shown in Fig.~1(b). 
The simple equation of motion is expressed as
\begin{equation}
M\frac{d^2z}{dt^2}=-\gamma \frac{dz}{dt}-F\cos(kz-\omega t),
\end{equation}
where $M$ is the mass of the particle, $\gamma$ is a damping constant, and $F$ is the strength of the wavy force. 
A new variable $z^{\prime}=z-\omega t/k$ satisfies
\begin{equation}
M\frac{d^2z^{\prime}}{dt^2}=-\gamma \frac{dz^{\prime}}{dt}-\frac{\omega \gamma}{k}-F\cos(kz^{\prime}).
\end{equation}
If $\gamma\omega/k\le F$, there is a steady solution of $z^{\prime}=\cos^{-1}[-\gamma\omega/(kF)]/k$. The solution corresponds to a steadily moving solution $z(t)=\omega t/k+\cos^{-1}[-\gamma\omega/(kF)]/k$. The particle motion is entrained to the wave motion. For $\gamma\omega/k\ge F$, there is no steady solution of $z^{\prime}$, that is, the particle motion delays from the traveling wave. 
We have performed direct numerical simulation for $M=k=1$, and $\omega=0.2$.
Figure 3(a) shows a steadily moving solution with the velocity $\omega/k$ at $F=0.22$, and a delayed motion at $F=0.18$.   Figure 3(b) shows the average velocity of the particle as a function of $F$ for $M=k=\gamma=1$, and $\omega=0.2$. There is an entrainment transition at $F=\gamma\omega/k$, above which the particle moves with the same velocity as the traveling wave.
\section{Peristaltic Motion Induced by Desynchronized Limit-Cycle Oscillators}
The electrophysiological rhythm in the small intestine is not synchronized. It is known that the frequency is higher near the stomach, and  the frequency profile has a stepwise structure~\cite{Davenport,Ermentrout}. 
  The desynchronization can be described by the phase oscillator model in purely discrete systems but the oscillation amplitude  becomes zero periodically at the desynchronization point for phase slips in a continuous system or in a discrete system with a large coupling constant. That is, the amplitude of oscillation is necessary to describe the desynchronization phenomena correctly. We use the Stuart-Landau equation to describe the desynchronized oscillation~\cite{Sakaguchi}.
The model equation is expressed as
\begin{equation}
\frac{dW_j}{dt}=(1+i\omega_{0j})W_j-(1+ic_2)|W_j|^2W_j+D(W_{j+1}-2W_j+W_{j-1}),\;{\rm for}\; 1\le j\le N,
\end{equation} 
where $W_j(t)=X_j(t)+iY_j(t)=A_j(t)e^{i\phi_j(t)}$ is a complex variable, and $\omega_{0j}$ is the natural frequency of the $j$th oscillator, which is assumed to be $\omega_{b}+\alpha j$. The natural frequency increases in space with a constant rate. The boundary conditions of $W_0=W_1$ and $W_{N+1}=W_N$ are imposed. The parameters $\omega_b,c_2$, and $D$ are fixed to be $\omega_b=0.5,c_2=0.5$, and $D=25$. The system size $N$ is assumed to be $N=400$. 

\begin{figure}[tbp]
\begin{center}
\includegraphics[height=4.cm]{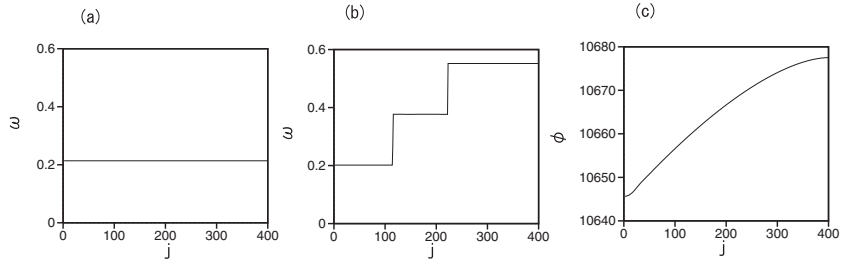}
\end{center}
\caption{Frequency profiles $\omega_j$ in Eq.~(11) at (a) $\alpha=0.0006$ and (c) 0.0015. (c) Phase profile $\phi_j$ at $\alpha=0.0006$.}
\label{fig5}
\end{figure}
\begin{figure}[tbp]
\begin{center}
\includegraphics[height=4.cm]{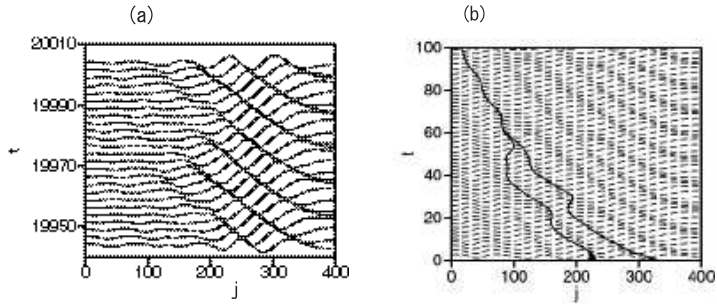}
\end{center}
\caption{(a) Time evolutions of $R_j$ in Eq.~(12) at $\alpha=0.0015$. (b) Amplitudes $\Delta R_j$ of deformation waves at $\alpha=0.0015$. (b) Time evolutions of $z_n(t)$ and $R_j(t)$ in Eqs.~(12) and (13) at  $\alpha=0.0015$ and $F=600$.}
\label{fig6}
\end{figure}
The average frequency in the coupled system is defined as $\omega_j=[\phi_j(t)-\phi_j(t_0)]/(t-t_0)$ for a sufficiently large time difference $t-t_0$. 
Figures 4(a),(b) show the average frequencies at (a) $\alpha=0.0006$ and (b) $\alpha=0.0015$.
At $\alpha=0.0006$, the whole oscillators are synchronized.
 At $\alpha=0.0015$, there are three entrained clusters. The boundaries of clusters locate at $j=115$ and 223.  
Figure 4(c) shows a phase profile $\phi_j$ at $\alpha=0.0006$.  The phase gradient $\partial \phi/\partial x$ denotes a local wavenumber. In the completely synchronized state at $\alpha=0.0006$, the local wavenumber decreases with $j$. 

We investigate a coupled system of Eq.~(4) as a simple model of radius oscillation of the small intestine, although actual motions of smooth muscles induced by electrophysiological signals are more complicated. 
The model equation is written as 
\begin{equation}
\frac{d^2R_j}{dt^2}=4K\sin^2(\Delta\theta/2)(R_j-R_0-R_1X_j)+d(R_{j+1}-2R_j+R_{j-1})-H\frac{dR_j}{dt}.
\end{equation} 
The radius oscillation is assumed to be induced by the oscillation of $X_j={\rm Re}\,W_j$ in Eq.~(11).  
Figure 5(a) shows the time evolution of $R_j$ at $\alpha=0.0015$. 
The other parameters are $\Delta\theta=2\pi/40, K=10,R_0=2,R_1=0.2,d=10$, and $H=0.05$.
Waves of radius oscillation move to the left, which corresponds to peristaltic motion.  There are three regions corresponding to the three frequency clusters shown in Fig.~4(b). The wave propagation is perturbed more largely. The amplitude of waves is large around $j=300$. 
 
Next, we investigate a simple model of a particle motion of mass $M$ driven by the deformation wave. We assume a simple model equation:
\begin{equation}
M\frac{d^2z}{dt^2}=-\gamma\frac{dz}{dt}-\frac{\partial}{\partial x}\left (\frac{F}{R^2(z)}\right ),
\end{equation}
where the tube radius $R(z)$ is calculated using Eq.~(12).  The deformation of the tube radius propagates as a wave as shown in Fig.~5(a). The term $F/R^2$ can be interpreted as a kind of pressure. $F$ denotes the strength of the force. The particle tends to be trapped in a region of large $R$ as shown in Fig.~1(b). We investigate whether particles are entrained to desynchronized traveling waves. 
 We have performed direct numerical simulations for 50 different initial conditions. the initial positions of $z(0)$ are set to be $220+0.5\times n$ for $n=1,2,\cdots,25$ and $320+0.5\times (n-25)$ for $n=26,27,\cdots,50$. 
Figure 5(b) shows time evolutions of $z_n(t)$ and $R_j(t)$ for $\alpha=0.0015$ and $F=600$. The other parameters are $\gamma=1,c_2=0.5,D=25$, $\omega_{b}=0.5$, $N=400$, $\Delta\theta=2\pi/40, K=10,R_0=2,R_1=0.2,d=10$, and $H=0.05$.  For this parameter, particles are scattered at the transition regions, but are successfully entrained to the wave, and move with the same velocity as the wave. 
However, the transportation of particles by the deformation waves becomes unsuccessful for $\alpha>0.0022$ owing to the desynchronization effect.
\section{Beating Motion Induced By Spatiotemporal Chaos} 
The Aliev-Panfilov model is a two-variable phenomenological model for the heart muscle~\cite{Aliev,Sakaguchi2}. The model equation is expressed as
\begin{eqnarray}
\frac{\partial e_{i,j}}{\partial t}&=&Ke_{i,j}(e_{i,j}-a)(e_{i,j}-1)-e_{i,j}r_{i,j}+D\Delta_d e_{i,j}, \nonumber\\
\frac{\partial g_{i,j}}{\partial t}&=&[e_{i,j}+\mu_1g_{i,j}/(\mu_2+e_{i,j})][-g_{i,j}-ke_{i,j}(e_{i,j}-b-1)].
\end{eqnarray}
Here, $e_{i,j}$ stands for the membrane potential and $g_{i,j}$ stands for the conductance of the inward current at the $(i,j)$ site. The parameter values $K,a,\epsilon,\mu_1,\mu_2$, and $b$ are evaluated on the basis of experiments on dogs as $K=8,\epsilon=0.01,\mu_1=0.11,\mu_2=0.3$, and $b=0.1$. The parameter $a$ is changed as a control parameter. $\Delta_d e_{i,j}$ denotes a discrete version of the Laplacian $\Delta_de_{i,j}=(e_{i+1,j}+e_{i-1,j}+e_{i,j+1}+e_{i,j-1}-4e_{i,j})$ in this paper. The uniform state $e=g=0$ is a stable solution. However, the system is excitable, and spiral waves appear under a certain stimulation.     
 
We consider a beating motion of a square sheet, which is expressed by a two-dimensional square lattice model coupled with springs. The equation of motion for the position vector ${\bf r}_{i,j}=(x_{i,j},y_{i,j})$ of the $(i,j)$ site is assumed as
\begin{equation}
m \frac{d^2{\bf r}_{i,j}}{dt^2}=-\gamma\frac{d{\bf r}_{i,j}}{dt}+\sum_{i^{\prime},j^{\prime}}K_{i^{\prime}.j^{\prime}}\{|{\bf r}_{i^{\prime},j^{\prime}}-{\bf r}_{i,j}|-q_{i^{\prime},j^{\prime}}[a_0-a_1(e_{i^{\prime},j^{\prime}}/2+e_{i,j}/2-e_0)]\}\frac{{\bf r}_{i^{\prime},j^{\prime}}-{\bf r}_{i,j}}{|{\bf r}_{i^{\prime},j^{\prime}}-{\bf r}_{i,j}|},
\end{equation}
where $m$ is mass and $\gamma$ is a damping constant, and the summation is taken for the nearest and next-nearest neighbors of $i,j$. $K_{i^{\prime},j^{\prime}}$ takes a value $K_1$ for the nearest neighbors, $K_2$ for the next-nearest neighbors, $a_0$ is the natural length of the spring, $q_{i^{\prime},j^{\prime}}=1$ for the nearest neighbors and $q_{i^{\prime},j^{\prime}}=\sqrt{2}$ for the 
next-nearest neighbors. We have assumed that the natural length of the spring between the two sites $(i,j)$ and $(i^{\prime},j^{\prime})$ changes as $a_0-a_1\{(e_{i,j}+e_{i^{\prime},j^{\prime}})/2-e_0\}$, which is proportional to the average value of the membrane potentials of the two sites.  
At the outer boundaries of the sheet, only interactions with inner neighbors are taken into account, as assumed in Eq.~(5). This is a simple model of a myocardial cell sheet. 
The oscillation of the expansion and contraction is experimentally observed in the cultivated myocardial cell sheet.

The natural length $a_0$ is assumed to be 1, and the initial conditions are $e_{i,j}=g_{i,j}=0$ and $x_{i,j}=i,y_{i,j}=j$. Firstly, a sinusoidal force of $0.05\sin(2\pi t/100)$ is added to the equation of $e_{i,j}$ in Eq.~(14) as an external stimulus. The membrane potential changes periodically with period $T=100$. The other parameters are $\gamma=0.5,m=0.002, K_1=1,K_2=0.25,a_0=1, a_1=0.3,e_0=0.314$, and $D=1$. The system size is $150\times 150$.  The sheet repeats expansion and contraction by the stimuli to the membrane potential. Figure 6(a) shows snapshots of the sheet at $t=727$ (contracted phase) and $t=778$ (expanded phase). Figure 6(b) shows the time evolution of the total area $S$ of the sheet. 
A strong mechanical beating appears. This type of strong beating is used for the pumping of blood flow in the heart. 
\begin{figure}[tbp]
\begin{center}
\includegraphics[height=4.cm]{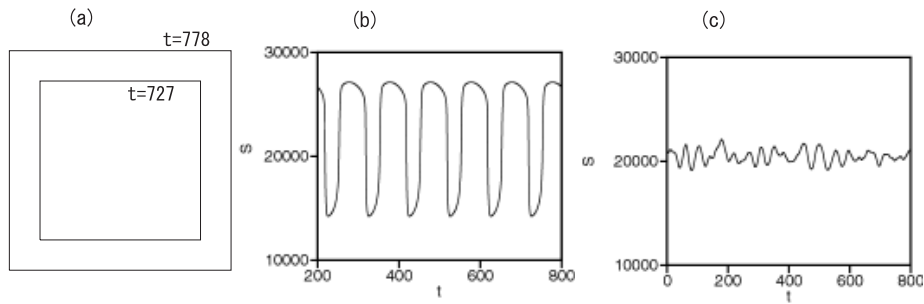}
\end{center}
\caption{(a) Snapshots of the sheet at $t=727$ and $t=778$. (b) Time evolution of the total area $S$ of the sheet by the regular pulsation. (c) Time evolution of the total area $S$ of the sheet by spiral chaos at $a_0=0.09$.}
\label{fig9}
\end{figure}

If a spiral pattern is set as an initial condition, regular rotating spiral waves are emitted from a spiral core at $a=0.13$. The external sinusoidal force is not added in this simulation. 
Even if there is no external sinusoidal force, mechanical motions are induced by the spiral waves of $e$.  
For a smaller $a$, spiral chaos appears.\cite{Sakaguchi2} 
Figure 6(c)  shows the time evolution of the total area $S$ at $a=0.09$. 
The outer shape of the sheet is irregularly deformed owing to the spatiotemporal chaos. 
At $a=0.09$, a chaotic fluctuation of $S(t)$ is observed. 
The temporal variation of total area is rather smaller than the case of the regular beating motion stimulated by a periodic force shown in Fig.~6(b). 
This result corresponds to the fact that strong heart beats disappear in ventricular fibrillation caused by spirals.    
\section{Summary}
We have studied several mechanical motions induced by spatially distributed limit-cycle oscillators.  Motivated by peristaltic motion in the small intestine, we have studied deformation waves of tube radius induced by desynchronized limit-cycle oscillators and particle motions driven by the deformation waves. 
We have shown that particles can be flown downward by the deformation waves, even if the entrainment to the wave cannot be attained, that is, the peristaltic motion works effectively for the transportation of food particles even if a few desynchronization clusters exist. However, the transportation stops when the desynchronization further proceeds. Next, we have studied the beating motions of a sheet induced by periodic stimuli, a regular spiral, and spiral chaos, which is motivated by the arrhythmia in the heart. We have reproduced that strong beating motions disappear when spirals appear, which is interpreted as a state of ventricular fibrillation.  Our models are very simple for the qualitative understanding of the coupling between the mechanical motion and the spatially distributed limit-cycle oscillators. We would like to construct more quantitative models incorporating physiological values as the next step.

\end{document}